\documentclass[aps,pre,twocolumn,superscriptaddress]{revtex4}

\usepackage{graphicx}
\usepackage{dcolumn}
\usepackage{bm}
\usepackage{amssymb}

\begin{document}


\title{Shape transformations of toroidal vesicles}

\author{Hiroshi Noguchi}
\email[]{noguchi@issp.u-tokyo.ac.jp}
\affiliation{
Institute for Solid State Physics, University of Tokyo,
 Kashiwa, Chiba 277-8581, Japan}
\author{Ai Sakashita}
\affiliation{Department of Physics, Ochanomizu University, 2-1-1 Otsuka, Bunkyo, Tokyo 112-8610, Japan}
\affiliation{
Institute for Solid State Physics, University of Tokyo,
 Kashiwa, Chiba 277-8581, Japan}
\author{Masayuki Imai}
\affiliation{Department of Physics, Faculty of Science, Tohoku University, 6-3 Aramaki, Aoba-ku, Sendai, Miyagi 980-8578, Japan}


\begin{abstract}
Morphologies of genus-1 and 2 toroidal vesicles are studied numerically by dynamically triangulated membrane models
and experimentally by confocal laser microscopy. 
Our simulation results reproduce shape transformations observed in our experiments well.
At large reduced volumes of the genus-1 vesicles,
obtained vesicle shapes
agree with the previous theoretical prediction, in which axisymmetric shapes are assumed: double-necked stomatocyte, discoidal toroid, and
circular toroid.
However, for small reduced volumes, it is revealed that a non-axisymmetric discoidal toroid 
and handled discocyte
exist in thermal equilibrium in the parameter range, in which the previous theory predicts axisymmetric discoidal shapes.
Polygonal toroidal vesicles and subsequent budding transitions are also found.
The entropy caused by shape fluctuations slightly modifies the stability of the vesicle shapes.
\end{abstract}


\maketitle

\section{Introduction}

Vesicles are closed bilayer membranes that show a 
wide variety of morphologies depending on the 
lipid architecture as well as their environment.
Since fluid lipid membranes are the main component of biomembranes,
lipid vesicles are considered as a simple model system of cells.
In particular, vesicle shapes with genus $g=0$
have been intensively investigated and are well understood  \cite{lipo95,lipo99,seif97,svet02,svet89,khal08,nogu09,saka12,saka14}.
For example, the shape of red blood cells, discocyte, can be formed by a lipid membrane without proteins.
In contrast to the genus-0 vesicles,
vesicles with nonzero genus have been much less explored.
In this paper, we focus on vesicles with $g=1$ and $g=2$.

In living cells, organelles exist in various shapes.
In some  organelles, lipidic necks or pores connect biomembranes
such that they have nonzero genus.
For example, the nuclear membrane and endoplasmic reticulum
are connected and together form complicated shapes.
The nucleus is wrapped by two bilayer membranes
 connected by many lipidic pores.
Thus, its shape is considered as a stomatocyte of a high-genus vesicle
connected with a tubular network.
It is important to understand how their topologies affect their morphologies.

Vesicle shapes are determined by the curvature energy and
the area difference $\Delta A$ of two monolayers of bilayer membranes 
with a constraint on the reduced volume $V^*$ \cite{lipo95,lipo99,seif97,svet02},
which is the volume relative to a spherical vesicle of the same surface area.
It is defined as $V^*=V/(4\pi/3){R_{\rm A}}^{3}$ with $R_{\rm A}=\sqrt{A/4\pi}$, where $V$ and $A$ are the vesicle volume and surface area, respectively.
Since transverse diffusion (flip--flop) of phospholipids between two monolayers is very slow,
the number difference $\Delta N_{\rm {lip}}$ of lipids between two monolayers can be conserved on a typical experimental time scale.
In the bilayer coupling (BC) model, the area difference $\Delta A$ is fixed as the preferred value
$\Delta A_0=\Delta N_{\rm {lip}}a_{\rm {lip}}$, where $a_{\rm {lip}}$ is the area per lipid in tensionless membranes.
In the area-difference-elasticity (ADE) model, 
a harmonic potential for the difference $\Delta A - \Delta A_0$ is added as a penalty for the deviation of the area difference.
For genus-0 vesicles, various observed morphologies can be reproduced well by both BC and ADE models \cite{lipo95,lipo99,seif97,svet02}: 
stomatocyte, discocyte, prolate, pear, pearl-necklace, and branched starfish-like shapes.
However, shape-transformation dynamics is better explained by the ADE model~\cite{saka12}.

The vesicle shapes with $g=1$ and $g=2$ were studied in the 1990s \cite{zhon90,seif91a,four92,juli93,juli93a,juli96,mich94,mich95,mich95a}.
For $g=1$, the phase diagrams of axisymmetric shapes were constructed for the BC, ADE, and spontaneous-curvature models
by J{\"u}licher et al.~\cite{juli93}.
They assumed axisymmetry of the vesicle shape and
the region of non-axisymmetric shapes is only estimated by a stability analysis of the axisymmetric shapes.
However, the stability is only examined with respect to special conformal transformations
and non-axisymmetric shapes were not directly explored.
Thus, the full phase diagram of genus-1 vesicles has not been completed.
For $g=2$, conformational degeneracy was found in the ground state at $V^*\gtrsim 0.7$,
where the vesicles can transform their shapes without changing their curvature energy with fixed $V^*$ and $\Delta A$~\cite{juli93a,juli96,mich95}.
In this paper, we revisit the phase diagram of $g=1$ using three-dimensional simulations 
and find non-axisymmetric thermal-equilibrium shapes 
in the region where axisymmetric shapes were assumed in Ref. \cite{juli93}.
In all of the previous theoretical studies on non-zero genus vesicle, the thermal fluctuations are neglected.
However, experimentally, neck diffusion of toroidal vesicles and  bending and length fluctuations of lipid tubes
were reported \cite{mich94,mich07,yama12}. 
We also investigate the effects of thermal fluctuations and 
compare our simulation results with shape transformations observed in our experiments for $g=1$ and $2$.

The simulation and experimental methods are provided in Sec. \ref{sec:method}.
In Sec. \ref{sec:g1},
vesicle shapes at $g=1$ are described.
First, the stable states for the curvature energy without the bilayer-coupling constraint or ADE energy
is explained as a starting point.
Subsequently the free-energy profiles are calculated for three values of the reduced volume.
This calculation clarifies discrete shape transitions from non-axisymmetric discoidal shapes
to circular toroids in the ADE model.
The simulation results are compared with experimental images.
The budding transitions are also discussed.
In Sec. \ref{sec:g2} the results for $g=2$ are described.
The summary and conclusions are given in Sec. \ref{sec:sum}.

\section{Materials and Methods}
\label{sec:method}

\subsection{Dynamically Triangulated Membrane Model}

We employ a dynamically triangulated surface model to describe a fluid membrane
\cite{nogu09,gomp97f,gomp04c}.
The vesicle consists of
$N_{\rm {mb}}$ vertices connected by bonds (tethers) to form a 
triangular network. 
The vertices have a hard-core excluded volume of diameter $\sigma_0$. 
The maximum length of the bond is $\sigma_1$.
In order to keep  the volume $V$ and surface area $A$ constant,
harmonic potentials 
$U_{\rm {V}}= (1/2)k_{\rm {V}}(V-V_{\rm 0})^2$ and
$U_{\rm {A}}= (1/2)k_{\rm {A}}(A-A_{\rm 0})^2$
are employed.
A Metropolis Monte Carlo (MC) method is used to move vertices.

The curvature energy of a single-component fluid vesicle is given by~\cite{canh70,helf73}
\begin{equation}
U_{\rm {cv}} =  \int  \frac{\kappa}{2}(C_1+C_2)^2   dA,
\label{eq:cv}
\end{equation}
where $C_1$ and $C_2$ are the principal curvatures at each point 
in the membrane. The coefficient $\kappa$ is the bending rigidity.
The spontaneous curvature vanishes when lipids are symmetrically distributed 
in both monolayers of the bilayer.
The integral over the Gaussian curvature $C_1C_2$ is omitted
because it is invariant for a fixed topology.

In the ADE model, the ADE energy $U_{\rm {ADE}}$ is added \cite{lipo95,lipo99,seif97,svet02,svet89}:
\begin{equation}
U_{\rm {ADE}} =  \frac{\pi k_{\rm {ade}}}{2Ah^2}(\Delta A - \Delta A_0)^2.
\label{eq:ade}
\end{equation}
The areas of the outer and inner monolayers of a bilayer vesicle
differ by $\Delta A= h \int (C_1+C_2) dA$,
where $h$ is the distance between the two monolayers.
The BC model can be considered as the ADE model with $k_{\rm {ade}}=\infty$.
The area differences are normalized by a spherical vesicle as
$\Delta a =\Delta A/8\pi h R_{\rm A}$ and $\Delta a_0 = \Delta A_0/8\pi h R_{\rm A}$
 to display our results.
For the spherical vesicle with $\Delta a_0=0$, $\Delta a =1$ and $U_{\rm {ADE}} =8\pi^2 k_{\rm {ade}}$.

The mean curvature at the $i$-th vertex is discretized as  \cite{itzy86,gomp04c,nogu05} 
\begin{equation}
(C_1+C_2){\bf n}_i = \frac{1}{\lambda_i}  
  \sum_{j(i)} \frac{\lambda_{i,j}{\bf r}_{i,j}}{r_{i,j}},
\end{equation}
where the sum over $j(i)$ is for the neighbors of the $i$-th vertex, which are
connected by bonds. The bond vector between the vertices $i$ and $j$ is
${\bf r}_{i,j}$, and $r_{i,j}=|{\bf r}_{i,j}|$. 
The length of a bond in the dual lattice is 
$\lambda_{i,j}=r_{i,j}[\cot(\theta_1)+\cot(\theta_2)]/2$.
The angles $\theta_1$ and $\theta_2$ are opposite to bond $ij$ in 
the two triangles sharing this bond,
  and $\lambda_i=0.25\sum_{j(i)} \lambda_{i,j}r_{i,j}$ is the area of the dual cell.
The normal vector ${\bf n}_i$ points from inside of the vesicle to outside.

The bonds are reconstructed
by flipping them to the diagonal of two adjacent triangles
using the Metropolis MC procedure.
Triangle formation of the bonds outside of the membrane surface is rejected
such that the minimum pore in vesicles consists of four bonds
[see the middle snapshot in Fig.~\ref{fig:min}(a)].
In the present simulations,
we use $N_{\rm {mb}}=1000$, $\sigma_1/\sigma_0=1.67$, $k_{\rm {V}}=4$, $k_{\rm {A}}=8$,
and $\kappa=20k_{\rm B}T$, 
where $k_{\rm B}T$ is the thermal energy.
The deviations of reduced volume $V^*$ from the specified values are less than $0.01$.
We primarily use $k_{\rm {ade}}^* = k_{\rm {ade}}/\kappa=1$,
which is a typical value for phospholipids~\cite{seif97,saka12}.
In the previous study~\cite{juli93}, the phase diagram of genus-$1$ vesicles was constructed
using this value.

The canonical MC simulations of the ADE model are performed with various parameter sets
for the potential $U=U_{\rm {cv}}+U_{\rm {V}}+U_{\rm {A}}+U_{\rm {ADE}}$.
To obtain the thermal equilibrium states,
one of the generalized ensemble MC methods \cite{okam04,berg03,nogu05} is employed for genus-1 vesicles.
Instead of the ADE potential $U_{\rm {ADE}}$, a weight potential $U_{\rm {w}}(\Delta A)$ is employed for a flat probability distribution over $\Delta A$.
Since the weight potential $U_{\rm w}(\Delta A)$ is not known a priori, it has to be estimated using an iterative procedure.
After long simulations, the canonical ensemble of the ADE model is obtained by
a re-weighting procedure \cite{ferr88}. 
In the case of the BC model, the canonical ensemble for the potential $U_0=U_{\rm {cv}}+U_{\rm {V}}+U_{\rm {A}}$ is calculated for a small bin of $\Delta a$ with a bin width of $0.0025$.
We perform annealing simulation to $T=0$
and the canonical MC simulations with $N_{\rm {mb}}=4000$  at several parameter sets to confirm the energy-minimum shapes
and the finite size effects of the triangulation, respectively.

\subsection{Experimental Method}

We prepared single-component vesicles from DOPC (1,2-dioleoyl-sn-glycero-3-phosphocholine, Avanti Polar Lipids) 
using the gentle hydration method with deionized water \cite{saka12}. 
TR-DHPE (Texas Red, 1,2-dihexadecanoyl-sn-glycero-3-phosphoethanolamine, Molecular Probes) was used as the dye ($1$\% mole ratio).
We kept vesicle suspensions at room temperature ($24$-$25^\circ$C) and observed them using a fast confocal laser microscope (Carl Zeiss, LSM 5Live). 
At this stage, most vesicles spontaneously formed either a spherical or tubular shape.
We observed vesicles with genus $g=1$ or $2$ by analyzing numerous microscopy images.

We observed shape transformations of the vesicles  of $g=1$ and $2$.
The intrinsic area difference $\Delta a_0$ is varied without changing the osmotic pressure.
In Ref. \cite{saka12},
we calculated $V$, $A$, and $\Delta a$ using the 3D images of genus-0 liposomes
and found that  during shape transformations,
$\Delta a$ is changed, whereas $V$ and $A$ are constant.
We concluded that small lipid reservoirs such as small lipid aggregates and bicelles are likely present on the membrane, and
the laser illumination of the microscope induces fusion into either monolayer of the lipid bilayer, 
which leads to changes in $\Delta a_0$.
We applied this method to toroidal vesicles here.
It is difficult 
 to measure 3D shapes of the observed toroidal vesicles owing to smallness of the liposomes or low contrast of the images.
It is not distinguishable only from the experiments
whether observed shapes are in thermal equilibrium or in metastable state.
We compare the experimental vesicle images and simulation snapshots
and determine that they are in  equilibrium or not.

\section{Results and Discussions}
\subsection{Genus-1 Toroidal Vesicles}
\label{sec:g1}

\subsubsection{Curvature Energy Model. }

\begin{figure}
\includegraphics{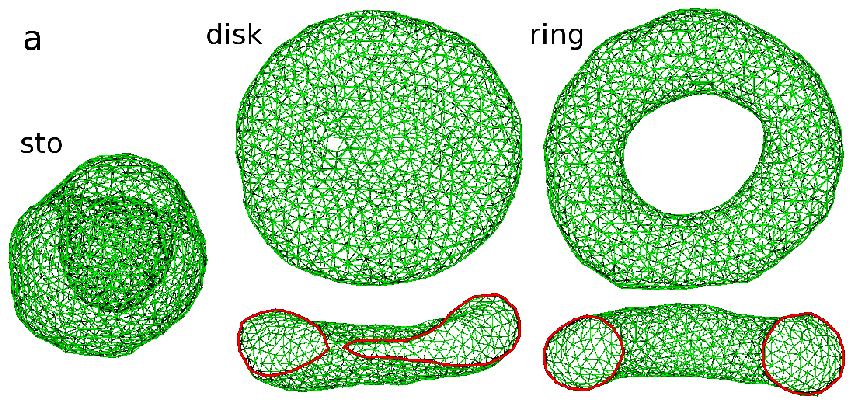}
\includegraphics{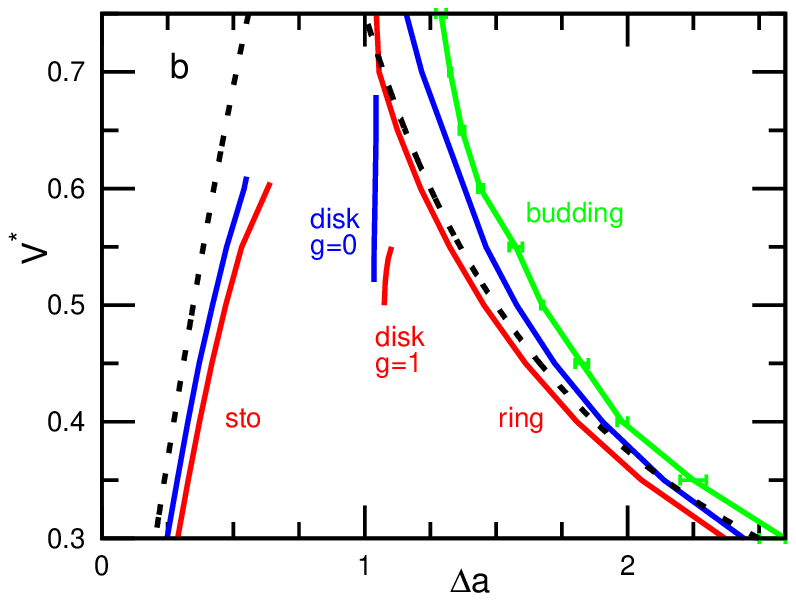}
\caption{
Stable and meta-stable shapes of vesicles in the curvature energy model.
(a) Snapshots of vesicles of genus $g=1$ at $V^*=0.5$.
The labels `sto,' 'disk,' and 'ring' represent stomatocyte, discocyte, and circular toroids, respectively.
The side views of half-cut snapshots are also shown for disk and ring shapes.
(b) Area difference $\Delta a$ of (meta-) stable shapes at (red) $g=1$ and (blue) $g=0$.
The green solid line represents  $\Delta a$ at the budding transition.
The dashed black lines are the results obtained using simple geometrical models for stomatocye and ring shapes.
}
\label{fig:min}
\end{figure}

First, we compare the (meta-) stable shapes of genus-1 vesicles and genus-0 vesicles without the ADE energy
or the BC constraint (see Fig.~\ref{fig:min}). 
The vesicles are simulated for the potential $U_0$.
We call this model the curvature energy model.
For the genus-0 vesicles, three axisymmetric shapes--stomatocyte, discocyte, and prolate-- are formed.
It is known that stomatocyte, discocyte, and prolate are the global-energy-minimum states 
for $0<V^*<0.59$, $0.59<V^*<0.65$, and $0.65<V^*<1$, respectively \cite{seif97}.
The shape transformations between these shapes are discrete transitions
such that these shapes exist as metastable states in wider ranges of $V^*$.
The free-energy-minimum states including metastable shapes are indicated by the blue lines in Fig.~\ref{fig:min}(b).

The genus-1 vesicles also have three energy-minimum shapes.
For stomatocyte and discocyte, an additional small neck or pore appears.
Instead of the prolate, a circular toroid is formed. 
Here, we abbreviate the shapes as 'sto,' 'disk,' and 'ring' for stomatocyte, discocyte, and circular toroid, respectively.
The discocyte exists in a narrower range of $V^*$ for $g=1$ than for $g=0$ [see Fig.~\ref{fig:min}(b)].
The previous theoretical study \cite{seif91a}
predicts these three types of the vesicles by minimizing the energy of the axisymmetric shapes.
The circular toroid can be approximated by the revolution of 
a circle as follows: 
\begin{equation}
{\bf r}(\theta, \phi)=( (r_1+r_2\cos\phi)\cos\theta,(r_1+r_2\cos\phi)\sin\theta, r_2\sin\phi ).
\label{eq:circ}
\end{equation}
In particular, the toroid with $r_1/r_2=\sqrt{2}$ is the Clifford torus
and has the lowest curvature energy $U_{\rm {Cl}}/\kappa=4\pi^2$
and $V^*_{\rm {CL}}=3/2^{5/4}\sqrt{\pi} \simeq 0.71$  \cite{zhon90}.
Note that thermal fluctuations give an additional curvature energy as 
$U_{\rm {cv}} \simeq U_{\rm {CL}} +  (N_{\rm {mb}}-1)k_{\rm B}T/2$ in the simulations 
because each bending mode is excited with $k_{\rm B}T/2$.
It was predicted that the ring vesicles are non-axisymmetric at $V^*>V^*_{\rm {CL}}$
but axisymmetric at $V^*\leq V^*_{\rm {CL}}$ \cite{seif91a}.
Our simulation results confirm their prediction for the ring.
Although the circular shapes at $V^*\leq V^*_{\rm {CL}}$  are not exactly axisymmetric,
its deviation is small and can be understood as thermal fluctuations around the circular toroids.
The area difference $\Delta a$ of the revolution expressed in Eq.~(\ref{eq:circ}) is shown
in the right black dashed line in Fig.~\ref{fig:min}(b).
Our simulation results agree well with this estimation.

\begin{figure}
\includegraphics{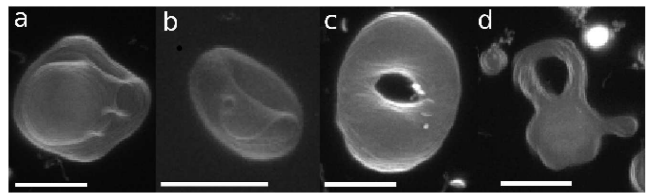}
\caption{
Microscopy images of genus-1 liposomes.
(a) Double-necked stomatycote. (b) Open stomatocyte.
(c) Ellipsoidal toroid. (d) Discocyte with a tubular handle and arm.
Scale bar: $10$ $\mu$m.
}
\label{fig:expg1}
\end{figure}

\begin{figure}
\includegraphics{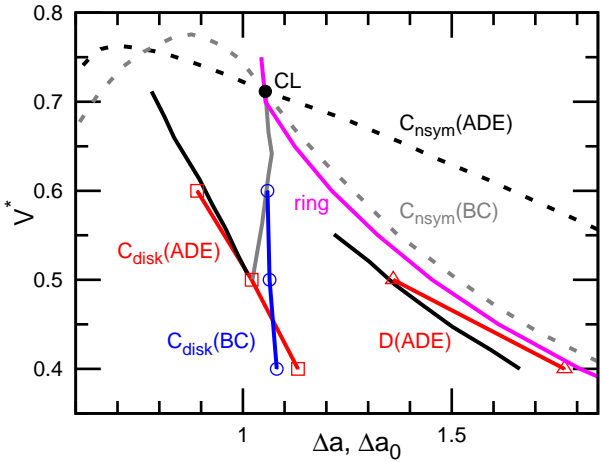}
\caption{
Shape transition lines of the genus-1 vesicles with the ADE and BC models.
The red and blue solid lines represent the shape transitions in ADE and BC models, respectively.
C$_{\rm disk}$ and D represent the continuous and discrete transitions between open stomatocyte and disk 
and between elongated disk and circular toroid, respectively.
The black and gray lines are ADE and BC boundaries extracted from J{\"u}licher's phase diagrams in Ref. \cite{juli93},
respectively.
The dashed lines represent phase boundaries between axisymmetric and non axisymmetric shapes (C$_{\rm nsym}$).
The filled circle represents the Clifford torus (CL). 
The magenta line represents the free-energy minimum states of the ring [the same as in Fig. \ref{fig:min}(b)]. 
}
\label{fig:pd}
\end{figure}

The axisymmetric double-necked stomatocyte (called a sickle-shaped toroid)
is predicted as the global-minimum state at low $V^*$ in the previous study \cite{seif91a}.
In our simulations, two necks (or pores in the top view) of the stomatocyte are not typically along the center axis 
[see the left snapshot in Fig.~\ref{fig:min}(a)].
Experimentally observed double-necked stomatocytes are also off-center [see Fig.~\ref{fig:expg1}(a)].
The narrow neck shape is nearly catenoid, and its mean curvature is negligibly small.
The energy difference between the axisymmetric and non-axisymmetric stomatocytes is small
unless two necks are close to each other.
The entropy of this diffusion is estimated as $S_{\rm pore} \simeq k_{\rm B}\ln(A/2d_{\rm {pore}}^2)$, where
$d_{\rm {pore}}$ is the diameter of the pores and 
$A/2$ is the average area of inner and outer spherical components of stomatocyte.
We obtain $S_{\rm pore} \simeq 5k_{\rm B}$ from $A=820\sigma^2$ and $d_{\rm sto} \simeq 2\sigma$ in our simulation.
This is not a large value but it seems to be sufficient to overcome the energy difference.
Thus, two necks can diffuse on the surface of the vesicle by thermal fluctuations.
Vesicle deformation coupled with neck diffusion is observed.
The membrane between two necks typically has large curvature [see Figs.~\ref{fig:min}(a) and \ref{fig:expg1}(a)].
The entropy of these shape fluctuations and pore shape fluctuations \cite{fara05}
may also increase stability of non-axisymmetric shapes of the stomatocyte.
The left black dashed line in Fig.~\ref{fig:min}(b) shows the values of $\Delta a$ 
of stomatocytes modeled using two spheres.
The shape deformation caused by thermal fluctuations slightly shift $\Delta a$ to larger values.

The pore in the discocyte is also off-center. We discuss this in detail in the next subsection.
In J{\"u}licher's phase diagrams \cite{juli93},
the equilibrium shapes are axisymmetric in large region including the stomatocyte and ring (see Fig.~\ref{fig:pd}).
We will show non-axisymmetric equilibrium shapes in this region for low $V^*$.

\begin{figure}
\includegraphics{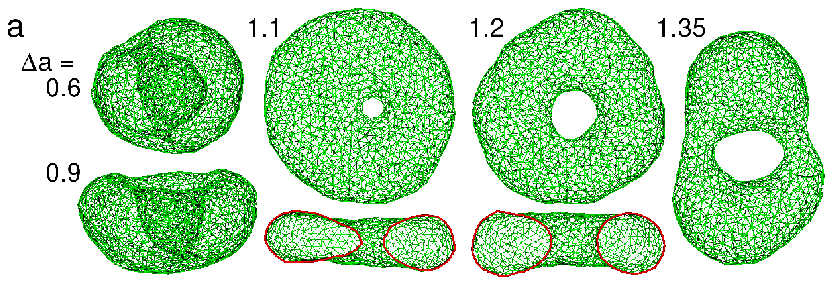}
\includegraphics{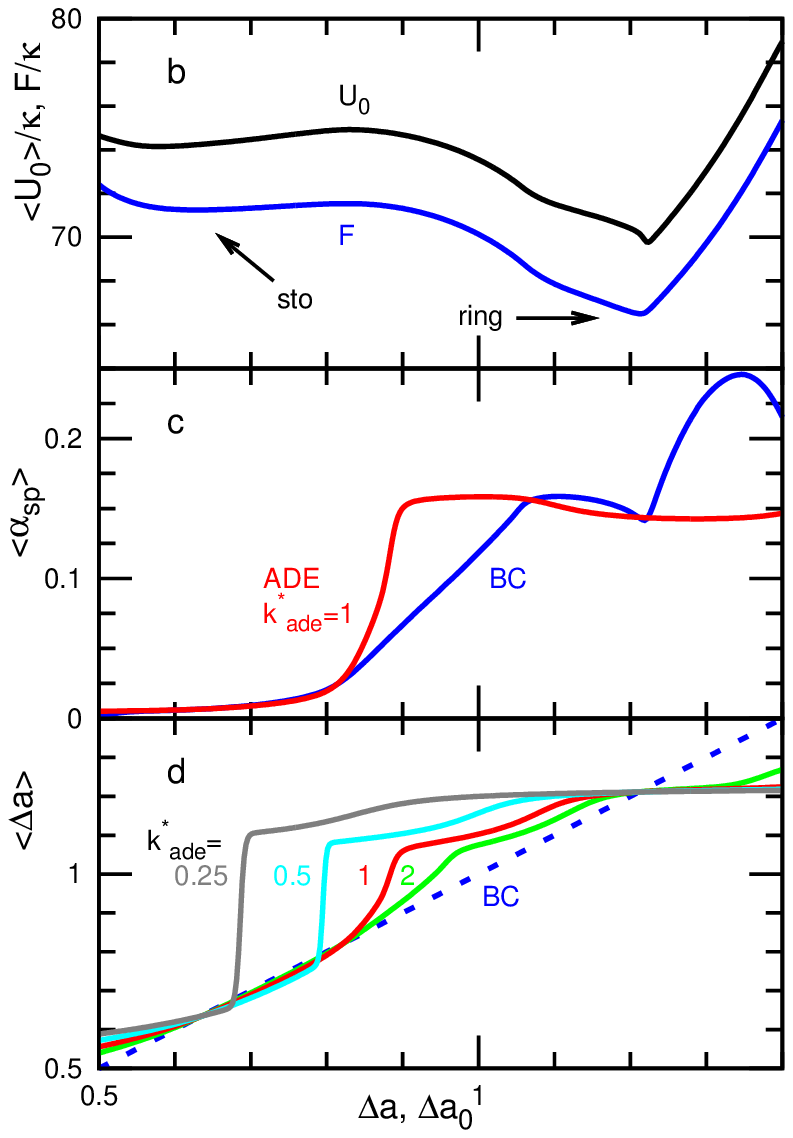}
\caption{
Dependence of shape on $\Delta a$ (BC) or $\Delta a_0$ (ADE) for  genus-1 vesicles at $V^*=0.6$.
(a) Snapshots at $\Delta a=0.6$, $0.9$, $1.1$, $1.2$, and $1.35$.
The side views of half-cut snapshots are also shown at $\Delta a=1.1$ and $1.2$.
(b) Free-energy profile $F$ and mean potential energy $\langle U_0 \rangle$ for the BC model.
(c) Mean asphericity $\langle \alpha_{\rm {sp}} \rangle$ for the BC and ADE models with $k_{\rm {ade}}^*=1$.
(d) Mean area difference $\langle \Delta a \rangle$.
The solid lines represent the ADE models with $k_{\rm {ade}}^*=0.25$, $0.5$, $1$, and $2$.
The dashed line represents the BC model ($\langle \Delta a \rangle = \Delta a$).
The error bars are smaller than the line thickness.
}
\label{fig:v6}
\end{figure}

\begin{figure}
\includegraphics{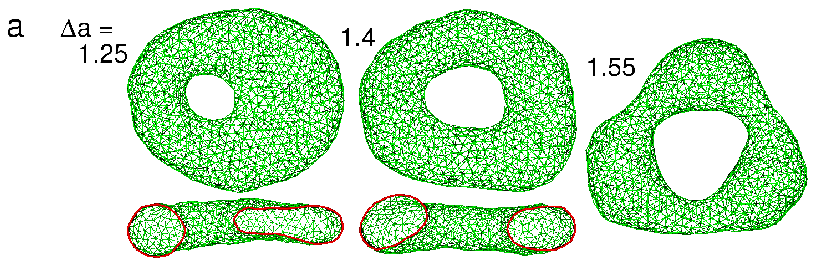}
\includegraphics{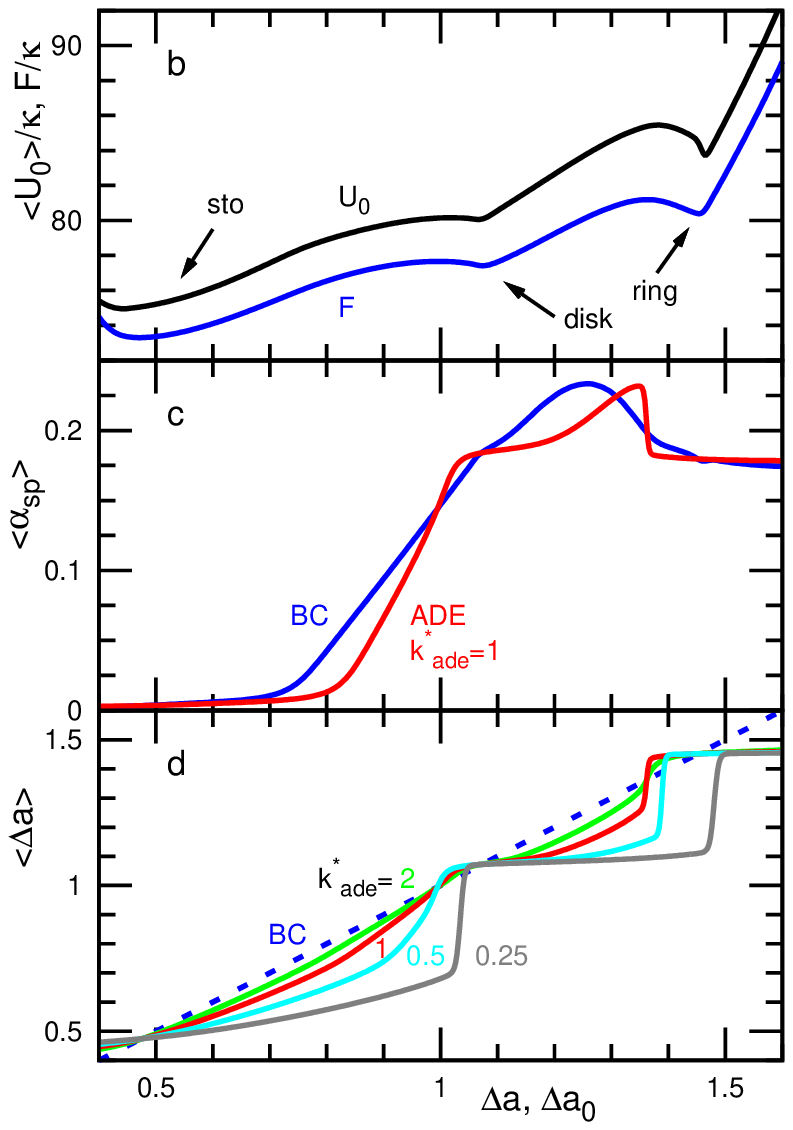}
\caption{
Dependence of shape on $\Delta a$ (BC) or $\Delta a_0$ (ADE) for genus-1 vesicles at $V^*=0.5$.
(a) Snapshots at $\Delta a=1.25$, $1.4$, and $1.55$.
The side views of half-cut snapshots are also shown at $\Delta a=1.25$ and $1.4$.
(b) Free-energy profile $F$ and mean potential energy $\langle U_0 \rangle$ for the BC model.
(c) Mean asphericity $\langle \alpha_{\rm {sp}} \rangle$ for the BC and ADE models with $k_{\rm {ade}}^*=1$.
(d) Mean area difference $\langle \Delta a \rangle$.
The data are presented in the same form as in Fig.~\ref{fig:v6}.
}
\label{fig:v5}
\end{figure}

\begin{figure}
\includegraphics{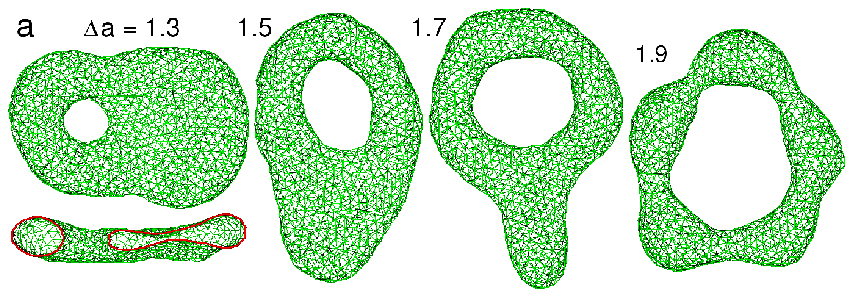}
\includegraphics{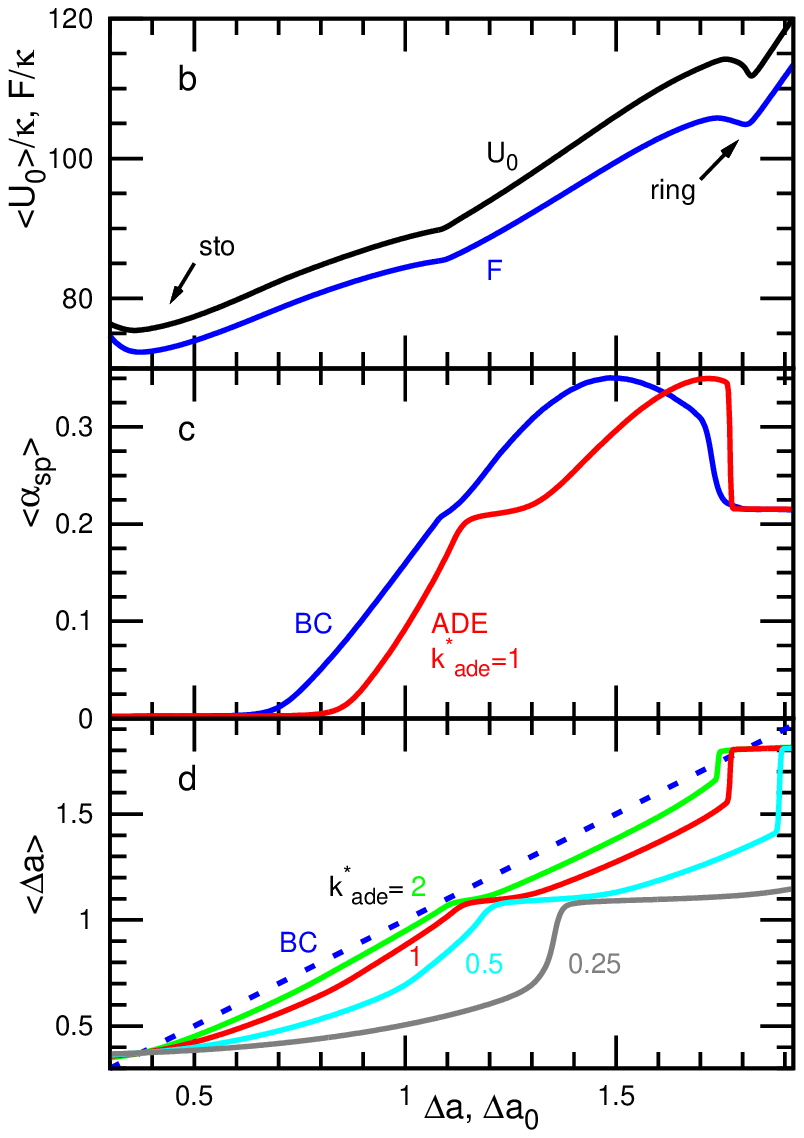}
\caption{
Dependence of shape on $\Delta a$ (BC) or $\Delta a_0$ (ADE) for genus-1 vesicles at $V^*=0.4$.
(a) Snapshots at $\Delta a=1.3$, $1.5$, $1.7$, and $1.9$.
The side view of the half-cut snapshot at $\Delta a=1.3$ is also shown.
(b) Free-energy profile $F$ and mean potential energy $\langle U_0 \rangle$ for the BC model.
(c) Mean asphericity $\langle \alpha_{\rm {sp}} \rangle$ for the BC and ADE models with $k_{\rm {ade}}^*=1$.
(d) Mean area difference $\langle \Delta a \rangle$.
The data are presented in the same form as in Fig.~\ref{fig:v6}.
}
\label{fig:v4}
\end{figure}

\subsubsection{BC and ADE Models. }

To clarify the phase behavior of the genus-1 toroidal vesicle,
the generalized ensemble MC simulations are performed at $V^*=0.4$, $0.5$, and $0.6$
(see Figs.~\ref{fig:pd}--\ref{fig:v4}).
The free-energy profiles $F$ are calculated using probability distributions $P_{\rm {cv}}(\Delta a)$
of the curvature energy model: $F(\Delta a)= -k_{\rm B}T\ln(P_{\rm {cv}}(\Delta a))+C$.
The constant $C$ is unknown. Thus, we cannot obtain the absolute values of $F$ but
can compare its relative values. In Figs.~\ref{fig:v6}--\ref{fig:v4}(b),
we shift $F$ to be close to $\langle U_0\rangle$ in order to make the comparison of two curves easier.
The difference between $\Delta a$ dependences of $F$ and $\langle U_0\rangle$ is 
caused by the entropy of membrane fluctuations.

To quantify the vesicle shapes, a shape parameter called asphericity, $\alpha_{\rm {sp}}$, is calculated.
It is defined as
 \cite{rudn86}
\begin{equation}
\alpha_{\rm {sp}} = \frac{({\lambda_1}-{\lambda_2})^2 + 
  ({\lambda_2}-{\lambda_3})^2+({\lambda_3}-{\lambda_1})^2}{2 (\lambda_1+\lambda_2+\lambda_3)^2},
\end{equation}
where ${\lambda_1} \leq {\lambda_2} \leq {\lambda_3}$ are the 
eigenvalues of the gyration tensor of the vesicle.
The asphericity is the degree of deviation from a spherical 
shape;  $\alpha_{\rm {sp}} = 0$ for spheres, $\alpha_{\rm {sp}}=1$ 
for thin rods, and $\alpha_{\rm {sp}}=0.25$ for thin disks.
The stomatocytes have  $\alpha_{\rm {sp}} \simeq 0$.
The disk and ring shapes have $\alpha_{\rm {sp}} \simeq 0.15$, $0.19$, and $0.21$ 
at $V^*=0.6$, $0.5$, and $0.4$, respectively.

For $V^*=0.6$ with increasing $\Delta a$,
a neck of the stomatocyte opens, and subsequently the stomatocyte transforms into a discocyte (see Fig.~\ref{fig:v6}).
As $\Delta a$ increases further, the pore in the discocyte becomes larger, 
and ultimately a circular toroid is formed.
As $\Delta a$ increases even further, the toroid elongates in one direction.
This elongated shape is also observed in our experiment [compare Figs. \ref{fig:expg1}(c) and \ref{fig:v6}(a)].
These transformations are continuous in both BC and ADE models with $k_{\rm {ade}}^*\geq 1$.
When the vesicle transforms from the discocyte to the open stomatocyte,
mirror symmetry breaks,
and the slops of $F(\Delta a)$ and the other quantities are changed.
This is a second-order type of the transition, but
the transition point is rounded by the finite energy increase.
Since the curvature energy is independent of the vesicle size,
the transition is not sharp even at $N_{\rm {mb}}=\infty$.
The area differences $\Delta a$ and $\Delta a_0$  at this transition point 
are estimated from the second derivative of  $\alpha_{\rm sp}$ curves in Fig. \ref{fig:v6}(c).
They agree with J{\"u}licher's results for both ADE and BC models
(see the curves denoted by C$_{\rm disk}$ in Fig.~\ref{fig:pd}).
The pore of the discocyte appears near the center of the disk.
As the vesicle is annealed to $T=0$, the pore moves to the center.
Thus, it is considered that the stable shape is the axisymmetric discocyte predicted in Ref. \cite{juli93}
and that the discocyte is slightly deformed by thermal fluctuations.
The discocyte does not have a minimum in the free-energy profile in Fig.~\ref{fig:v6}(b),
and no clear transitions are observed between the discocyte and circular toroid.
The neck in the open stomatocyte can also be off-center.
Such a non-axisymmetric open stomatocyte is experimentally observed, as shown in Fig.~\ref{fig:expg1}(b).

\begin{figure}
\includegraphics{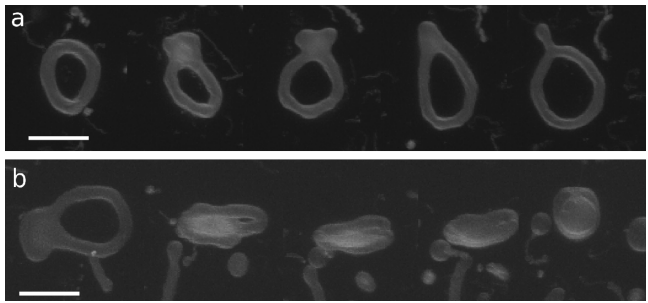}
\caption{
Time-sequential microscopy images of genus-1 liposomes.
(a) Transformation from a circular toroid to racket shape via a handled discocyte.
From left to right: $t=0$, $160$, $180$, $188$, and $192$ s.
(b) Transformation from a handled discocyte to stomatocyte via a discocyte with a small pore and open stomatocyte.
From left to right: $t=0$, $62$, $80$, $91$, and $131$ s.
Scale bar: $10$ $\mu$m.
}
\label{fig:snapv3s}
\end{figure}

At $V^*=0.5$, the vesicle transforms from a stomatocyte to a circular toroid via a discocyte
[see Figs.~\ref{fig:min}(a) and \ref{fig:v5}].
The difference from $V^*=0.6$ is that the discocyte is at a minimum of $F(\Delta a)$
and non-axisymmetric elongated discocytes appear between two energy minima of the discocyte and ring.
In the stable discocyte [the middle snapshot in Fig.~\ref{fig:min}(a)], 
the pore is off-center and stays on the edge of the dimple.
As the vesicle is annealed to $T=0$, the pore remain on the edge.
We  also confirmed the formation of these non-axisymmetric discocytes at $N_{\rm {mb}}=4000$. 
Michalet and Bensimon experimentally observed this off-center discocyte \cite{mich95a}.
They reported that it is an energy minimum state by using the energy minimization, 
but they did not clarify whether it is the global- or local- minimum state~\cite{mich95a}.
Our simulation revealed that the off-center discocyte is the global-minimum state in both BC and ADE models.
In the axisymmmetric discocyte,
the pore opens in the almost flat membranes at the middle of the dimples.
As the pore approaches the edge of the dimples,
the curvature of the edge is partially reduced.
Similar pore formations on the edge of the highly curved structures
are obtained in membrane-fusion intermediates.
The fusion pore opens at the edge of hemifusion diaphragm \cite{nogu02c,nogu12}
and at the side of a stalk neck connecting two bilayer membranes \cite{nogu01b,mull11}.

\begin{figure}
\includegraphics{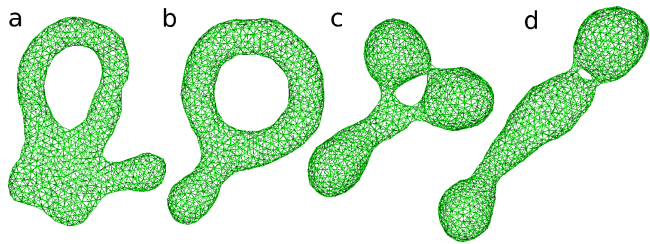}
\caption{
Snapshots of genus-1 vesicles.
(a) $(V^*, \Delta a_0)=(0.3,2.2)$.
(b) $(V^*, \Delta a_0)=(0.4,1.9)$.
(c) $(V^*, \Delta a_0)=(0.5,2.2)$.
(d) $(V^*, \Delta a_0)=(0.5,2.2)$.
}
\label{fig:snapg1}
\end{figure}

As $\Delta a$ increases at $V^*=0.5$, the pore in the disocyte is expanded at this off-center position,
and the vesicle becomes an elliptic discocyte 
[see the left snapshot in Fig.~\ref{fig:v5}(a) and the peak at $\Delta a=1.25$ in Fig.~\ref{fig:v5}(c)].
In the BC model, the vesicle then becomes a circular toroid via an elongated circular toroid 
[see the middle snapshot in Fig.~\ref{fig:v5}(a)].
In contrast, in the ADE model with $k_{\rm {ade}}^*=1$, 
the vesicle exhibits a discrete transition from an elliptic discocyte 
to a circular toroid  at $\Delta a_0=1.36$.
This transition point agrees with the prediction in Ref. \cite{juli93} 
(see Fig. \ref{fig:pd}),
despite the fact that only axisymmetric shapes are considered in Ref. \cite{juli93}.
This good agreement might be due to the small energy difference between the axisymmetric discocyte and elliptic discocyte.

\begin{figure}
\includegraphics{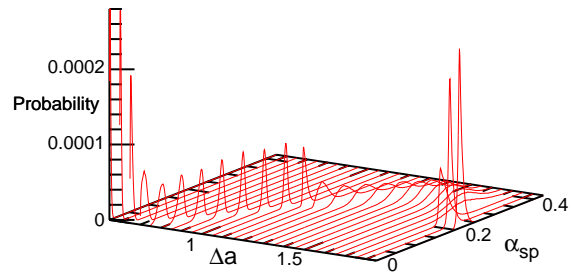}
\caption{
Probability distribution of asphericity $\alpha_{\rm {sp}}$  for genus-1 vesicles in the BC model at $V^*=0.4$.
}
\label{fig:asp}
\end{figure}

At $V^*=0.4$, the vesicle forms pronounced non-axisymmetric shapes 
between the off-center circular discocyte and circular toroid (see Fig.~\ref{fig:v4}).
The pore is not in the dimples of the discocyte but outside of the discotye, which is similar to a handle,
as shown in the right snapshot in Fig.~\ref{fig:v4}(a).
We also observed this shape experimentally (see the left images in Fig.~\ref{fig:snapv3s}).
Such handled discocyte has not been reported previously.
With decreasing $V^*$, the elliptic discocyte becomes more elongated,
and subsequently the large flat parts form dimples.
As $\Delta a$ increases, the pore becomes larger and the discoidal part becomes narrower.
When $\Delta a$ is further increased in the BC model, the vesicles form 
a tubular arm similar to the grip of a racket [see Fig.~\ref{fig:v4}(a)]
and subsequently becomes a circular toroid.
In the ADE model, the racket-shaped vesicle is skipped in the phase diagram,
and a discrete transition from the elongated handled discocyte of $\Delta a=1.25$ to the circular toroid of $\Delta a=1.81$
occurs at $\Delta a_0=1.77$ [see Fig.~\ref{fig:v4}(c)].
The transition point is slightly shifted from the previous prediction
for the transition from axisymmetric disk and ring \cite{juli93}
(see Fig. \ref{fig:pd}),
because of a large deviation of the discocyte from the axisymmetric shape.
The racket-shaped vesicle exists as a local-minimum state at larger values of $\Delta a_0$
[see Fig.~\ref{fig:snapg1}(b)].
In the experiment, a handled discocyte with a tubular arm is observed [see Fig.~\ref{fig:expg1}(d)].
A similar shape is obtained as a local energy-minimum state in the ADE model [see Fig.~\ref{fig:snapg1}(a)].
A circular toroid connecting two tubular arms is also obtained (data not shown).
At low $V^*$, tubular arms often remain once they are formed such that several local-minimum states appear.

Figure~\ref{fig:snapv3s} shows time-sequential microscopy images of shape transformations of genus-1 vesicles.
These transformations from a handled discocyte to a racket shape and stomatocyte are well explained by gradual changes of $\Delta a_0$ at $V^*=0.3$:
from $\Delta a_0=2.2$ to $2.6$ [Fig.~\ref{fig:snapv3s}(a)] and from $\Delta a_0=2.3$ to $0.6$ 
[Fig.~\ref{fig:snapv3s}(b)], respectively.
Movies of corresponding simulations are provided in ESI (Movies 1 and 2).

The entropy of the shape fluctuations slightly modifies the phase behavior.
The area difference $\Delta a$ at the minimum free energy
is smaller than that at the minimum of the mean potential energy $\langle U_0 \rangle$ by  $0.01$  at $V^*=0.5$.
The entropy of the shape fluctuations is large for the elongated discocyte
because the difference of the curves $\langle U_0 \rangle$ and $F$ are large at $\Delta a\simeq 1.4$ and $1.75$
 at $V^*=0.5$ and $0.4$, respectively.
The fluctuations of asphericity $\alpha_{\rm {sp}}$ are also large in these regions, and
 $\alpha_{\rm {sp}}$ exhibits broad distributions, as shown in Fig.~\ref{fig:asp}.

We have shown the discrete transition between elongated discocyte and circular toroid at $V^*=0.5$ and $0.4$ for the ADE model with  $k_{\rm {ade}}^*=1$.
The transformation between stomatocyte and discocyte becomes a discrete transition at lower values of $k_{\rm {ade}}^*$ [see Figs. \ref{fig:v6}(d) and \ref{fig:v5}(d)].
With decreasing  $V^*$, a lower value of $k_{\rm {ade}}^*$ is needed to obtain the discrete transition.
Thus, the values of  $k_{\rm {ade}}^*$ may be estimated by  systematically observing the transformation dynamics under changes in $\Delta a_0$ or $V^*$.

As $\Delta a$ increases further from the circular toroid, 
the toroid deforms into ellipsoid, triangle, and pentagon at $V^*=0.6$, $0.5$, and $0.4$, respectively.
Thus, a higher undulation mode becomes unstable and grows at lower values of $V^*$.
When the vesicle is approximated as the circular revolution expressed in Eq.~(\ref{eq:circ}),
the length ratio of two circumferences are $r_1/r_2=1.99$, $2.86$, and $4.48$ at $V^*=0.6$, $0.5$, and $0.4$, respectively.
Therefore, the unstabilized mode is determined by the wavelength of $2\pi r_2$.
When $r_1/r_2\to\infty$, the toroid is approximated as a cylinder of radius $r_2$. 
It is known that for cylindrical membranes of spontaneous curvature $C_0$
the undulation mode of wavelength $2\pi r_2$ becomes unstable at $C_0r_2=1$ \cite{zhon89}.
Our simulation results show that this relation holds for finite values of  $r_1/r_2$.

\begin{figure}
\includegraphics{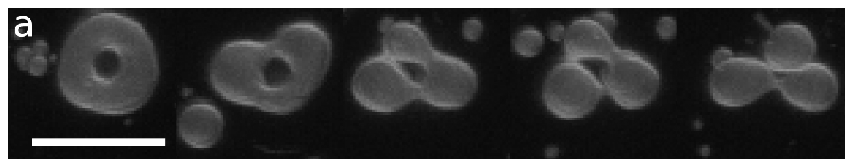}
\includegraphics{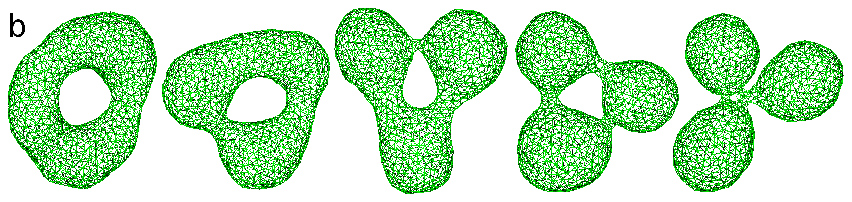}
\caption{
Shape transformations of a genus-1 vesicle at $V^*=0.55$.
(a)  Time-sequential microscopy images of liposomes.
From left to right: $t=0$, $45$, $95$, $97$, and $144$ s.
Scale bar: $10$ $\mu$m.
(b) Sequential snapshots of the triangulated-membrane simulation.
Intrinsic area difference $\Delta a_0$ is gradually increased with $d\Delta a_0/dt= 1\times 10^{-8}$,
where $t$ represents MC steps.
From left to right: $\Delta a_0=1.73$, $1.8$, $1.87$, $1.97$, and $2.35$.
}
\label{fig:snapv55s}
\end{figure}

\subsubsection{Budding. }

The toroidal vesicles exhibit a budding transition in a manner similar to genus-0 vesicles.
The area difference $\Delta a$ at the budding transition for $g=1$ is shown as a green line in Fig.~\ref{fig:min}(b).
This line is obtained though canonical MC simulations of the ADE model.
The budding is a discrete transition, and the error bars show the hysteresis regions,
where budded or unbudded shapes are obtained on the basis of initial vesicle conformations.
The intrinsic area difference $\Delta a_0$ at the transition points are larger than  $\Delta a$: 
$\Delta a_0=2.6\pm 0.1$ and $1.29\pm 0.02$ at $V^*=0.3$ and $0.75$, respectively.
An example of the budding transition is shown in Fig.~\ref{fig:snapv55s}.
Our simulation reproduces the dynamics of the liposome very well.

Since the budded compartments are divided by small necks,
large free-energy barriers can exist between meta-stable and stable states.
Thus, it is difficult to identify the most stable state.
The snapshots in Figs.~\ref{fig:snapg1}(c) and (d) show two free-energy-minimum states at $(V^*, \Delta a_0)=(0.5,2.2)$.
The triangular and straight shapes are typically obtained as $\Delta a_0$ increases and as $V^*$ decreases, respectively.
These two shapes have almost identical potential energy with $\langle U \rangle/\kappa= 120$.
As $\Delta a_0$ gradually decreases from $\Delta a_0=2.2$, 
the straight shape is retained better
than the triangular budded shape at lower values of $\Delta a_0$.
A bud connected by two necks appears to be more robust than one connected by a single neck.
To calculate the free energies using a generalized ensemble method, order parameters to connect these budded states 
and circular toroid are required. However, typical shape parameters such as $\alpha_{\rm {sp}}$
are not suitable, since the budding occurs in local regions of the vesicle.

\begin{figure}
\includegraphics{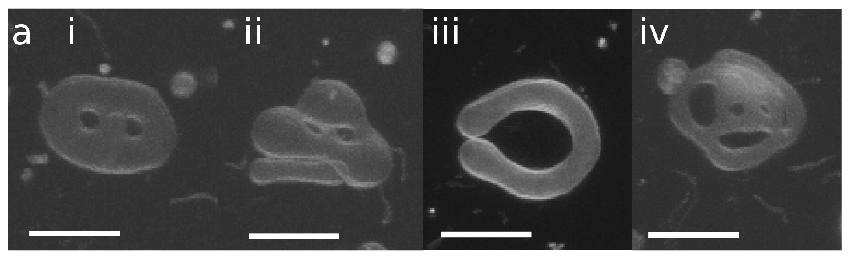}
\includegraphics{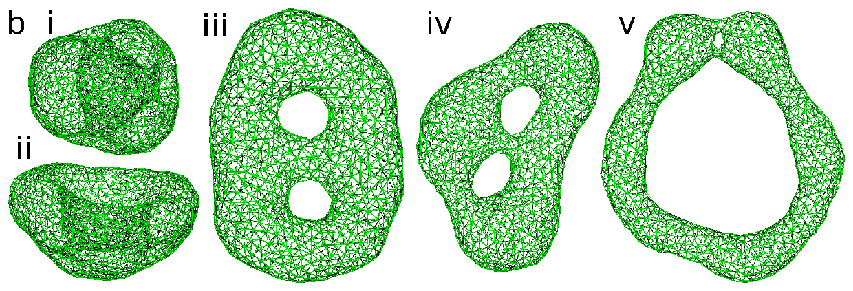}
\caption{
(a) Microscopy images of liposomes of (i--iii) $g=2$ and (iv) $g=5$.
Scale bar: $10$ $\mu$m.
(b) Snapshots for the toroidal vesicle of $g=2$.
(i) $(V^*, \Delta a_0)=(0.5,0.8)$.
(ii) $(V^*, \Delta a_0)=(0.5,1)$.
(iii) $(V^*, \Delta a_0)=(0.5,1.5)$.
(iv) $(V^*, \Delta a_0)=(0.5,1.8)$.
(v) $(V^*, \Delta a_0)=(0.35,2.7)$.
}
\label{fig:expg2}
\end{figure}

\subsection{Genus-2 Toroidal Vesicles}
\label{sec:g2}

\begin{figure}
\includegraphics{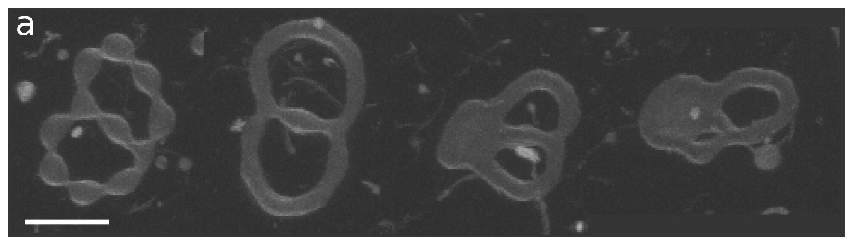}
\includegraphics{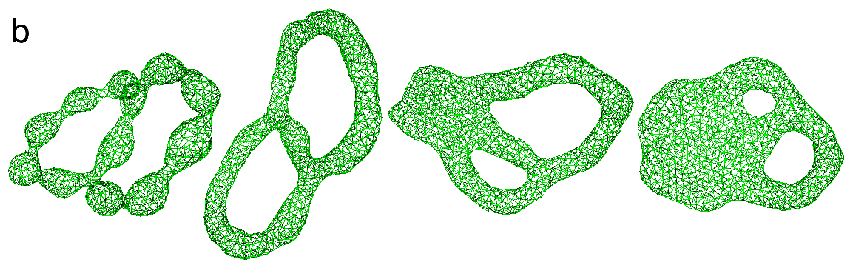}
\caption{
Shape transformations of a genus-2 vesicle at $V^*=0.25$.
(a)  Time-sequential microscopy images of liposomes.
From left to right: $t=0$, $84$, $197$, and $244$ s.
Scale bar: $10$ $\mu$m.
(b) Sequential snapshots of the triangulated-membrane simulation.
Intrinsic area difference $\Delta a_0$ is gradually decreased with $d\Delta a_0/dt= 2\times 10^{-8}$.
From left to right: $\Delta a_0=4$, $3.46$, $2.6$, and $2$.
}
\label{fig:snapv25s}
\end{figure}

The genus-2 vesicles form closed and open stomatocytes, discocytes, and budded shapes
(see Figs.~\ref{fig:expg2} and \ref{fig:snapv25s}).
These shapes are similar to those of genus-1 vesicles described in the previous section.
A significant difference is that no axisymmetric shapes exist for $g=2$.
Two pores cannot align at the center of the discocyte.

A discocyte with two handles is observed experimentally.
The time-sequential shape transformation from a budded toroid to handled discocyte 
is well reproduced by our simulation [compare Figs.~\ref{fig:snapv25s}(a) and (b)].

The size of two pores in the vesicles can be different.
An extreme example is shown in Figs.~\ref{fig:expg2}(a,iii) and (b,v).
The ends of a tubular vesicle are connected by two necks.
This shape is metastable and obtained in the region slightly below the budding transition.
At $g>2$, more complicated shapes are expected.
Figure~\ref{fig:expg2}(a,iv) shows an example of genus-5 liposomes.

\section{Conclusions}
\label{sec:sum}

Various morphologies of the genus-1 and 2 toroidal vesicles are clarified numerically and experimentally.
For $g=1$, 
new stable non-axisymmetric shapes are revealed from the free-energy profile calculation. 
As $\Delta a$  increases at $V^*=0.5$, a circular discocyte with an off-center pore transforms into
an elliptic discocyte and then into a circular toroid.
At $V^*=0.4$, between the circular discocyte and circular toroid, 
the discocyte with a handle and racket shape are formed.
The discrete transition occurs between the elliptic discocyte and circular toroid in the ADE model.
The handled discocytes are also observed for $g=2$.
As increasing $\Delta a$ further, the formation of polygonal toroidal vesicles and 
subsequential budding are also found.
Number of polygon edges are increased with decreasing $V^*$ and
are understood by the bending instability.
Our simulation results agree with our experimental observation very well. 
The following shapes are experimentally observed for the fist time:
Stomatocytes and racket-shaped vesicle for genus-1 and 
handled discocytes for genus-1 and 2.

As the genus of vesicles increases, the regions of non-axisymmetric shapes are expanded in the phase diagram.
For $g=1$, the non-axisymmetric discocytes are stable
in the region where the axisymmetric discocytes are formed for $g=0$.
For $g=2$, all of obtained shapes are non-axisymmetric.
Although the phase diagram of the genus-2 vesicles is investigated on the basis of symmetry analyses in Refs. \cite{juli93a,juli96},
thermal fluctuations are neglected.
A hexagonal array of pores is observed in polymersomes \cite{halu02}.
Shape transformations of the vesicles of genus 2 and higher under thermal fluctuations
are an interesting problem for further studies.

\begin{acknowledgments}
The authors would like to thank
 G. Gompper for informative discussions.
This work was partially supported by a Grant-in-Aid for Scientific Research on Innovative Areas 
"Fluctuation \& Structure" (No. 25103010) from 
the Ministry of Education, Culture, Sports, Science, and Technology of Japan.
The numerical calculations were partly
carried out on SGI Altix ICE 8400EX 
at ISSP Supercomputer Center, University of Tokyo. 
\end{acknowledgments}


\begin{thebibliography}{37}
\expandafter\ifx\csname natexlab\endcsname\relax\def\natexlab#1{#1}\fi
\expandafter\ifx\csname bibnamefont\endcsname\relax
  \def\bibnamefont#1{#1}\fi
\expandafter\ifx\csname bibfnamefont\endcsname\relax
  \def\bibfnamefont#1{#1}\fi
\expandafter\ifx\csname citenamefont\endcsname\relax
  \def\citenamefont#1{#1}\fi
\expandafter\ifx\csname url\endcsname\relax
  \def\url#1{\texttt{#1}}\fi
\expandafter\ifx\csname urlprefix\endcsname\relax\def\urlprefix{URL }\fi
\providecommand{\bibinfo}[2]{#2}
\providecommand{\eprint}[2][]{\url{#2}}

\bibitem[{\citenamefont{Lipowsky and Sackmann}(1995)}]{lipo95}
\bibinfo{editor}{\bibfnamefont{R.}~\bibnamefont{Lipowsky}} \bibnamefont{and}
  \bibinfo{editor}{\bibfnamefont{E.}~\bibnamefont{Sackmann}}, eds.,
  \emph{\bibinfo{title}{Structure and Dynamics of Membranes}}
  (\bibinfo{publisher}{Elsevier Science}, \bibinfo{address}{Amsterdam},
  \bibinfo{year}{1995}).

\bibitem[{\citenamefont{Lipowsky}(1999)}]{lipo99}
\bibinfo{author}{\bibfnamefont{R.}~\bibnamefont{Lipowsky}}, in
  \emph{\bibinfo{booktitle}{Stastical mechanics of biocomplexity}}, edited by
  \bibinfo{editor}{\bibfnamefont{D.}~\bibnamefont{Reguera}},
  \bibinfo{editor}{\bibfnamefont{J.~M.} \bibnamefont{Rubi}}, \bibnamefont{and}
  \bibinfo{editor}{\bibfnamefont{J.~M.~B.} \bibnamefont{Vilar}}
  (\bibinfo{publisher}{Springer}, \bibinfo{address}{Berlin},
  \bibinfo{year}{1999}), vol. \bibinfo{volume}{527} of
  \emph{\bibinfo{series}{Lecture Notes in Physics}}, pp.
  \bibinfo{pages}{1--23}.

\bibitem[{\citenamefont{Seifert}(1997)}]{seif97}
\bibinfo{author}{\bibfnamefont{U.}~\bibnamefont{Seifert}},
  \bibinfo{journal}{Adv.\ Phys.} \textbf{\bibinfo{volume}{46}},
  \bibinfo{pages}{13} (\bibinfo{year}{1997}).

\bibitem[{\citenamefont{Svetina and \v{Z}ek\v{s}}(2002)}]{svet02}
\bibinfo{author}{\bibfnamefont{S.}~\bibnamefont{Svetina}} \bibnamefont{and}
  \bibinfo{author}{\bibfnamefont{B.}~\bibnamefont{\v{Z}ek\v{s}}},
  \bibinfo{journal}{Anat. Rec.} \textbf{\bibinfo{volume}{268}},
  \bibinfo{pages}{215} (\bibinfo{year}{2002}).

\bibitem[{\citenamefont{Svetina and \v{Z}ek\v{s}}(1989)}]{svet89}
\bibinfo{author}{\bibfnamefont{S.}~\bibnamefont{Svetina}} \bibnamefont{and}
  \bibinfo{author}{\bibfnamefont{B.}~\bibnamefont{\v{Z}ek\v{s}}},
  \bibinfo{journal}{Euro. Biophys. J.} \textbf{\bibinfo{volume}{17}},
  \bibinfo{pages}{101} (\bibinfo{year}{1989}).

\bibitem[{\citenamefont{Khalifat et~al.}(2008)\citenamefont{Khalifat, Puff,
  Bonneau, Fournier, and Angelova}}]{khal08}
\bibinfo{author}{\bibfnamefont{N.}~\bibnamefont{Khalifat}},
  \bibinfo{author}{\bibfnamefont{N.}~\bibnamefont{Puff}},
  \bibinfo{author}{\bibfnamefont{S.}~\bibnamefont{Bonneau}},
  \bibinfo{author}{\bibfnamefont{J.-B.} \bibnamefont{Fournier}},
  \bibnamefont{and} \bibinfo{author}{\bibfnamefont{M.~I.}
  \bibnamefont{Angelova}}, \bibinfo{journal}{Biophys. J.}
  \textbf{\bibinfo{volume}{95}}, \bibinfo{pages}{4924 } (\bibinfo{year}{2008}).

\bibitem[{\citenamefont{Noguchi}(2009)}]{nogu09}
\bibinfo{author}{\bibfnamefont{H.}~\bibnamefont{Noguchi}},
  \bibinfo{journal}{J.\ Phys.\ Soc.\ Jpn.} \textbf{\bibinfo{volume}{78}},
  \bibinfo{pages}{041007} (\bibinfo{year}{2009}).

\bibitem[{\citenamefont{Sakashita et~al.}(2012)\citenamefont{Sakashita,
  Urakami, Ziherl, and Imai}}]{saka12}
\bibinfo{author}{\bibfnamefont{A.}~\bibnamefont{Sakashita}},
  \bibinfo{author}{\bibfnamefont{N.}~\bibnamefont{Urakami}},
  \bibinfo{author}{\bibfnamefont{P.}~\bibnamefont{Ziherl}}, \bibnamefont{and}
  \bibinfo{author}{\bibfnamefont{M.}~\bibnamefont{Imai}},
  \bibinfo{journal}{Soft\ Matter} \textbf{\bibinfo{volume}{8}},
  \bibinfo{pages}{8569} (\bibinfo{year}{2012}).

\bibitem[{\citenamefont{Sakashita et~al.}(2014)\citenamefont{Sakashita, Imai,
  and Noguchi}}]{saka14}
\bibinfo{author}{\bibfnamefont{A.}~\bibnamefont{Sakashita}},
  \bibinfo{author}{\bibfnamefont{M.}~\bibnamefont{Imai}}, \bibnamefont{and}
  \bibinfo{author}{\bibfnamefont{H.}~\bibnamefont{Noguchi}},
  \bibinfo{journal}{Phys. Rev. E} \textbf{\bibinfo{volume}{89}},
  \bibinfo{pages}{040701(R)} (\bibinfo{year}{2014}).

\bibitem[{\citenamefont{Ou-Yang}(1990)}]{zhon90}
\bibinfo{author}{\bibfnamefont{Z.~C.} \bibnamefont{Ou-Yang}},
  \bibinfo{journal}{Phys. Rev. A} \textbf{\bibinfo{volume}{41}},
  \bibinfo{pages}{4517} (\bibinfo{year}{1990}).

\bibitem[{\citenamefont{Seifert}(1991)}]{seif91a}
\bibinfo{author}{\bibfnamefont{U.}~\bibnamefont{Seifert}},
  \bibinfo{journal}{Phys. Rev. Lett.} \textbf{\bibinfo{volume}{66}},
  \bibinfo{pages}{2404} (\bibinfo{year}{1991}).

\bibitem[{\citenamefont{Fourcade et~al.}(1992)\citenamefont{Fourcade, Mutz, and
  Bensimon}}]{four92}
\bibinfo{author}{\bibfnamefont{B.}~\bibnamefont{Fourcade}},
  \bibinfo{author}{\bibfnamefont{M.}~\bibnamefont{Mutz}}, \bibnamefont{and}
  \bibinfo{author}{\bibfnamefont{D.}~\bibnamefont{Bensimon}},
  \bibinfo{journal}{Phys. Rev. Lett.} \textbf{\bibinfo{volume}{68}},
  \bibinfo{pages}{2551} (\bibinfo{year}{1992}).

\bibitem[{\citenamefont{J{\"u}licher
  et~al.}(1993{\natexlab{a}})\citenamefont{J{\"u}licher, Seifert, and
  Lipowsky}}]{juli93}
\bibinfo{author}{\bibfnamefont{F.}~\bibnamefont{J{\"u}licher}},
  \bibinfo{author}{\bibfnamefont{U.}~\bibnamefont{Seifert}}, \bibnamefont{and}
  \bibinfo{author}{\bibfnamefont{R.}~\bibnamefont{Lipowsky}},
  \bibinfo{journal}{J.\ Phys.\ II\ (France)} \textbf{\bibinfo{volume}{3}},
  \bibinfo{pages}{1681} (\bibinfo{year}{1993}{\natexlab{a}}).

\bibitem[{\citenamefont{J{\"u}licher
  et~al.}(1993{\natexlab{b}})\citenamefont{J{\"u}licher, Seifert, and
  Lipowsky}}]{juli93a}
\bibinfo{author}{\bibfnamefont{F.}~\bibnamefont{J{\"u}licher}},
  \bibinfo{author}{\bibfnamefont{U.}~\bibnamefont{Seifert}}, \bibnamefont{and}
  \bibinfo{author}{\bibfnamefont{R.}~\bibnamefont{Lipowsky}},
  \bibinfo{journal}{Phys. Rev. Lett.} \textbf{\bibinfo{volume}{71}},
  \bibinfo{pages}{452} (\bibinfo{year}{1993}{\natexlab{b}}).

\bibitem[{\citenamefont{J{\"u}licher}(1996)}]{juli96}
\bibinfo{author}{\bibfnamefont{F.}~\bibnamefont{J{\"u}licher}},
  \bibinfo{journal}{J.\ Phys.\ II\ (France)} \textbf{\bibinfo{volume}{6}},
  \bibinfo{pages}{1797} (\bibinfo{year}{1996}).

\bibitem[{\citenamefont{Michalet et~al.}(1994)\citenamefont{Michalet, Bensimon,
  and Fourcade}}]{mich94}
\bibinfo{author}{\bibfnamefont{X.}~\bibnamefont{Michalet}},
  \bibinfo{author}{\bibfnamefont{D.}~\bibnamefont{Bensimon}}, \bibnamefont{and}
  \bibinfo{author}{\bibfnamefont{B.}~\bibnamefont{Fourcade}},
  \bibinfo{journal}{Phys. Rev. Lett.} \textbf{\bibinfo{volume}{72}},
  \bibinfo{pages}{168} (\bibinfo{year}{1994}).

\bibitem[{\citenamefont{Michalet and Bensimon}(1995{\natexlab{a}})}]{mich95}
\bibinfo{author}{\bibfnamefont{X.}~\bibnamefont{Michalet}} \bibnamefont{and}
  \bibinfo{author}{\bibfnamefont{D.}~\bibnamefont{Bensimon}},
  \bibinfo{journal}{Science} \textbf{\bibinfo{volume}{269}},
  \bibinfo{pages}{666} (\bibinfo{year}{1995}{\natexlab{a}}).

\bibitem[{\citenamefont{Michalet and Bensimon}(1995{\natexlab{b}})}]{mich95a}
\bibinfo{author}{\bibfnamefont{X.}~\bibnamefont{Michalet}} \bibnamefont{and}
  \bibinfo{author}{\bibfnamefont{D.}~\bibnamefont{Bensimon}},
  \bibinfo{journal}{J.\ Phys.\ II\ (France)} \textbf{\bibinfo{volume}{5}},
  \bibinfo{pages}{263} (\bibinfo{year}{1995}{\natexlab{b}}).

\bibitem[{\citenamefont{Michalet}(2007)}]{mich07}
\bibinfo{author}{\bibfnamefont{X.}~\bibnamefont{Michalet}},
  \bibinfo{journal}{Phys. Rev. E} \textbf{\bibinfo{volume}{76}},
  \bibinfo{pages}{02914} (\bibinfo{year}{2007}).

\bibitem[{\citenamefont{Yamamoto and Ichikawa}(2012)}]{yama12}
\bibinfo{author}{\bibfnamefont{A.}~\bibnamefont{Yamamoto}} \bibnamefont{and}
  \bibinfo{author}{\bibfnamefont{M.}~\bibnamefont{Ichikawa}},
  \bibinfo{journal}{Phys. Rev. E} \textbf{\bibinfo{volume}{86}},
  \bibinfo{pages}{061905} (\bibinfo{year}{2012}).

\bibitem[{\citenamefont{Gompper and Kroll}(1997)}]{gomp97f}
\bibinfo{author}{\bibfnamefont{G.}~\bibnamefont{Gompper}} \bibnamefont{and}
  \bibinfo{author}{\bibfnamefont{D.~M.} \bibnamefont{Kroll}},
  \bibinfo{journal}{J. Phys. Condens. Matter} \textbf{\bibinfo{volume}{9}},
  \bibinfo{pages}{8795} (\bibinfo{year}{1997}).

\bibitem[{\citenamefont{Gompper and Kroll}(2004)}]{gomp04c}
\bibinfo{author}{\bibfnamefont{G.}~\bibnamefont{Gompper}} \bibnamefont{and}
  \bibinfo{author}{\bibfnamefont{D.~M.} \bibnamefont{Kroll}}, in
  \emph{\bibinfo{booktitle}{Statistical Mechanics of Membranes and Surfaces}},
  edited by \bibinfo{editor}{\bibfnamefont{D.~R.} \bibnamefont{Nelson}},
  \bibinfo{editor}{\bibfnamefont{T.}~\bibnamefont{Piran}}, \bibnamefont{and}
  \bibinfo{editor}{\bibfnamefont{S.}~\bibnamefont{Weinberg}}
  (\bibinfo{publisher}{World Scientific}, \bibinfo{address}{Singapore},
  \bibinfo{year}{2004}), \bibinfo{edition}{2nd} ed.

\bibitem[{\citenamefont{Canham}(1970)}]{canh70}
\bibinfo{author}{\bibfnamefont{P.~B.} \bibnamefont{Canham}},
  \bibinfo{journal}{J. Theor. Biol.} \textbf{\bibinfo{volume}{26}},
  \bibinfo{pages}{61} (\bibinfo{year}{1970}).

\bibitem[{\citenamefont{Helfrich}(1973)}]{helf73}
\bibinfo{author}{\bibfnamefont{W.}~\bibnamefont{Helfrich}},
  \bibinfo{journal}{Z.\ Naturforsch} \textbf{\bibinfo{volume}{28c}},
  \bibinfo{pages}{693} (\bibinfo{year}{1973}).

\bibitem[{\citenamefont{Itzykson}(1986)}]{itzy86}
\bibinfo{author}{\bibfnamefont{C.}~\bibnamefont{Itzykson}}, in
  \emph{\bibinfo{booktitle}{Proceedings of the GIFT seminar, Jaca 85}}, edited
  by \bibinfo{editor}{\bibfnamefont{J.}~\bibnamefont{Abad}},
  \bibinfo{editor}{\bibfnamefont{M.}~\bibnamefont{Asorey}}, \bibnamefont{and}
  \bibinfo{editor}{\bibfnamefont{A.}~\bibnamefont{Cruz}}
  (\bibinfo{publisher}{World Scientific}, \bibinfo{address}{Singapore},
  \bibinfo{year}{1986}).

\bibitem[{\citenamefont{Noguchi and Gompper}(2005)}]{nogu05}
\bibinfo{author}{\bibfnamefont{H.}~\bibnamefont{Noguchi}} \bibnamefont{and}
  \bibinfo{author}{\bibfnamefont{G.}~\bibnamefont{Gompper}},
  \bibinfo{journal}{Phys.\ Rev.\ E} \textbf{\bibinfo{volume}{72}},
  \bibinfo{pages}{011901} (\bibinfo{year}{2005}).

\bibitem[{\citenamefont{Okamoto}(2004)}]{okam04}
\bibinfo{author}{\bibfnamefont{Y.}~\bibnamefont{Okamoto}}, \bibinfo{journal}{J.
  Mol. Graph. Model.} \textbf{\bibinfo{volume}{22}}, \bibinfo{pages}{425}
  (\bibinfo{year}{2004}).

\bibitem[{\citenamefont{Berg et~al.}(2003)\citenamefont{Berg, Noguchi, and
  Okamoto}}]{berg03}
\bibinfo{author}{\bibfnamefont{B.~A.} \bibnamefont{Berg}},
  \bibinfo{author}{\bibfnamefont{H.}~\bibnamefont{Noguchi}}, \bibnamefont{and}
  \bibinfo{author}{\bibfnamefont{Y.}~\bibnamefont{Okamoto}},
  \bibinfo{journal}{Phys.\ Rev.\ E} \textbf{\bibinfo{volume}{68}},
  \bibinfo{pages}{036126} (\bibinfo{year}{2003}).

\bibitem[{\citenamefont{Ferrenberg and Swendsen}(1988)}]{ferr88}
\bibinfo{author}{\bibfnamefont{A.~M.} \bibnamefont{Ferrenberg}}
  \bibnamefont{and} \bibinfo{author}{\bibfnamefont{R.~H.}
  \bibnamefont{Swendsen}}, \bibinfo{journal}{Phys.\ Rev.\ Lett.}
  \textbf{\bibinfo{volume}{61}}, \bibinfo{pages}{2635} (\bibinfo{year}{1988}).

\bibitem[{\citenamefont{Farago and Santangelo}(2005)}]{fara05}
\bibinfo{author}{\bibfnamefont{O.}~\bibnamefont{Farago}} \bibnamefont{and}
  \bibinfo{author}{\bibfnamefont{C.~D.} \bibnamefont{Santangelo}},
  \bibinfo{journal}{J.\ Chem.\ Phys.} \textbf{\bibinfo{volume}{122}},
  \bibinfo{pages}{044901} (\bibinfo{year}{2005}).

\bibitem[{\citenamefont{Rudnick and Gaspari}(1986)}]{rudn86}
\bibinfo{author}{\bibfnamefont{J.}~\bibnamefont{Rudnick}} \bibnamefont{and}
  \bibinfo{author}{\bibfnamefont{G.}~\bibnamefont{Gaspari}},
  \bibinfo{journal}{J. Phys.\ A:\ Math.\ Gen.} \textbf{\bibinfo{volume}{19}},
  \bibinfo{pages}{L191} (\bibinfo{year}{1986}).

\bibitem[{\citenamefont{Noguchi}(2002)}]{nogu02c}
\bibinfo{author}{\bibfnamefont{H.}~\bibnamefont{Noguchi}},
  \bibinfo{journal}{J.\ Chem.\ Phys.} \textbf{\bibinfo{volume}{117}},
  \bibinfo{pages}{8130} (\bibinfo{year}{2002}).

\bibitem[{\citenamefont{Noguchi}(2012)}]{nogu12}
\bibinfo{author}{\bibfnamefont{H.}~\bibnamefont{Noguchi}},
  \bibinfo{journal}{Soft Matter} \textbf{\bibinfo{volume}{8}},
  \bibinfo{pages}{3146} (\bibinfo{year}{2012}).

\bibitem[{\citenamefont{Noguchi and Takasu}(2001)}]{nogu01b}
\bibinfo{author}{\bibfnamefont{H.}~\bibnamefont{Noguchi}} \bibnamefont{and}
  \bibinfo{author}{\bibfnamefont{M.}~\bibnamefont{Takasu}},
  \bibinfo{journal}{J.\ Chem.\ Phys.} \textbf{\bibinfo{volume}{115}},
  \bibinfo{pages}{9547} (\bibinfo{year}{2001}).

\bibitem[{\citenamefont{M{\"u}ller and Schick}(2011)}]{mull11}
\bibinfo{author}{\bibfnamefont{M.}~\bibnamefont{M{\"u}ller}} \bibnamefont{and}
  \bibinfo{author}{\bibfnamefont{M.}~\bibnamefont{Schick}},
  \bibinfo{journal}{Curr.\ Top.\ Membr.} \textbf{\bibinfo{volume}{68}},
  \bibinfo{pages}{295} (\bibinfo{year}{2011}).

\bibitem[{\citenamefont{Ou-Yang and Helfrich}(1989)}]{zhon89}
\bibinfo{author}{\bibfnamefont{Z.~C.} \bibnamefont{Ou-Yang}} \bibnamefont{and}
  \bibinfo{author}{\bibfnamefont{W.}~\bibnamefont{Helfrich}},
  \bibinfo{journal}{Phys. Rev. A} \textbf{\bibinfo{volume}{39}},
  \bibinfo{pages}{5280} (\bibinfo{year}{1989}).

\bibitem[{\citenamefont{Haluska et~al.}(2002)\citenamefont{Haluska,
  G{\'o}{\'z}d{\'z}, D{\"o}bereiner, F{\"o}rster, and Gompper}}]{halu02}
\bibinfo{author}{\bibfnamefont{C.~K.} \bibnamefont{Haluska}},
  \bibinfo{author}{\bibfnamefont{W.~T.} \bibnamefont{G{\'o}{\'z}d{\'z}}},
  \bibinfo{author}{\bibfnamefont{H.-G.} \bibnamefont{D{\"o}bereiner}},
  \bibinfo{author}{\bibfnamefont{S.}~\bibnamefont{F{\"o}rster}},
  \bibnamefont{and} \bibinfo{author}{\bibfnamefont{G.}~\bibnamefont{Gompper}},
  \bibinfo{journal}{Phys. Rev. Lett.} \textbf{\bibinfo{volume}{89}},
  \bibinfo{pages}{238302} (\bibinfo{year}{2002}).

\end{thebibliography}
\end{document}